\documentclass[twocolumn,showpacs,preprintnumbers,amsmath,amssymb,superscriptaddress]{revtex4}

\usepackage{graphicx}
\usepackage{amssymb}
\usepackage{times}
\usepackage{amsmath}
\usepackage{dcolumn}
\usepackage{bm}
\usepackage{colordvi}

\begin{document}

\title{Measuring photon anti-bunching from continuous variable
  sideband squeezing}

\author{Nicolai B. Grosse} \affiliation{Quantum Optics Group,
  Department of Physics, Faculty of Science, The Australian National
  University, ACT 0200, Australia}

\author{Thomas Symul} \affiliation{Quantum Optics Group, Department of
  Physics, Faculty of Science, The Australian National University, ACT
  0200, Australia}

\author{Magdalena Stobi\'nska} \affiliation{Instytut Fizyki
  Teoretycznej, Uniwersytet Warszawski, 00-681 Warszawa, Poland.}

\author{Timothy C. Ralph} \affiliation{Department of Physics,
  University of Queensland, St Lucia, Australia.}

\author{Ping Koy Lam} \affiliation{Quantum Optics Group, Department of
  Physics, Faculty of Science, The Australian National University, ACT
  0200, Australia}

\begin{abstract}
  We present a technique for measuring the second-order coherence
  function $g^{(2)}(\tau)$ of light using a Hanbury-Brown Twiss
  intensity interferometer modified for homodyne detection. The
  experiment was performed entirely in the continuous variable regime
  at the sideband frequency of a bright carrier field. We used the
  setup to characterize $g^{(2)}(\tau)$ for thermal and coherent
  states, and investigated its immunity to optical loss. We measured
  $g^{(2)}(\tau)$ of a displaced squeezed state, and found a best
  anti-bunching statistic of $g^{(2)}(0) = 0.11 \pm 0.18$.
\end{abstract}

\date{\today}
\maketitle

Fifty years ago, Hanbury-Brown and Twiss (HBT) first demonstrated an
optical intensity interferometer \cite{HBT}.  Since then, HBT
interferometry has been applied to diverse areas such as condensed
matter physics, atomic physics, and quantum optics
\cite{Condensed1,Atomic,Walls&Milburn}; and has become a powerful
measurement technique in astronomy, and high-energy particle physics
\cite{Astronomy,Particles}.  From a historical perspective, HBT
reported correlations in the intensity measured at two locations, from
light emitted by a thermal source. The effect was interpreted as being
either a manifestation of classical wave theory, or due to {\it
  bunching} in the arrival time of photons. Such correlations were
generalized to {\it n}th-order by Glauber in a comprehensive quantum
theory of optical coherence \cite{Glauber}, with the second-order
coherence $g^{(2)}$ corresponding to the measurement made with a HBT
interferometer. Curiously, the theory predicted that certain states of
light would exhibit a photon {\it anti-bunching} effect, which is the
tendency for photons to arrive apart from one another.  This is a
non-classical phenomenon which violates the Schwartz inequality
\cite{Walls&Milburn}.  Photon anti-bunching has been observed in
resonance fluorescence, conditioned measurements of parametrically
down-converted light, pulsed parametric amplification, quantum dots,
and trapped single atoms/molecules
\cite{Kimble,RarityAndNogueira,Koashi,QDots,SinglePhoton}.  Recent
experiments have probed the spatial and temporal second-order
coherence functions of atomic species in BEC, and atom lasers
\cite{Schellekens,Oettl}.  All of these experiments have relied upon
the ability to detect individual particles in a time-resolved
measurement.

\begin{figure}
\center{\includegraphics[width=1\linewidth]{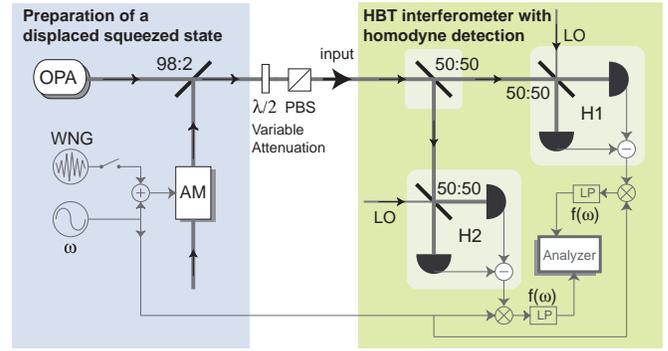}}
\caption{(Color online) Schematic of the experimental setup. OPA
  optical parametric amplifier, $\lambda /2$ half-wave plate, PBS
  polarizing beam-splitter, x:y beamsplitter with transmission x,
  H1/H2 homodyne detectors, AM amplitude modulator, $\omega$ function
  generator, WNG white noise generator, $\otimes$ mixer, LP low pass
  filter.}
\end{figure} 

In this Letter we apply a technique for measuring the second-order
coherence of optical fields, that complements previous studies and
provides a link between discrete-variable (DV) and continuous-variable
(CV) quantum optics. Our scheme is based on the HBT interferometer,
but uses homodyne detection in each arm, to make CV measurements of
the quadrature amplitudes over a range of sideband frequencies. The
second-order coherence $g^{(2)}$ is then constructed from the set of
four permutations of the time-averaged correlations between the
amplitude/phase quadratures.  At no point is it necessary to make
time-resolved detections of single-photons \cite{McAlisterAndWebb}.
Homodyne detection offers the advantage of high bandwidth, and
excellent immunity to extraneous optical modes. We used the scheme to
measure the temporal second-order coherence function $g^{(2)}(\tau)$
of a displaced squeezed state.

In contrast to most CV experiments involving squeezed light, is the
realization that weaker squeezed states can exhibit a greater
anti-bunching effect. Some properties of displaced squeezed states in
the context of second-order coherence have been investigated before
\cite{Koashi,StolerAndMahran}.  Such states can exhibit behavior
ranging from photon anti-bunching to super-bunching, provided that the
state is sufficiently pure, and the squeezing weak.  Using our
modified HBT interferometer, and exploiting the high stability and low
optical loss of our experimental setup, we were able to prepare and
measure displaced squeezed states that clearly demonstrated photon
anti-bunching. In addition, we investigated the immunity of
second-order coherence measurements to optical attenuation.

\begin{figure}
\center{\includegraphics[width=1\linewidth]{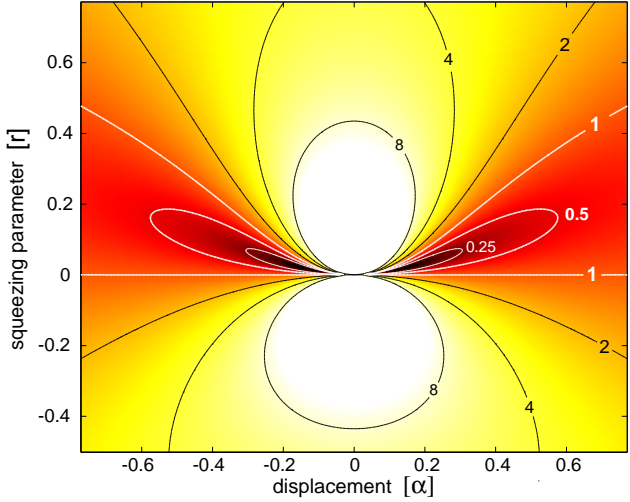}}
\caption{(Color online) The $g^{(2)}(0)$ parameter space for a displaced
  squeezed state with squeezing $r$ and displacement $\alpha$.
  Regions exhibiting photon anti-bunching are marked progressively
  darker, with contours giving the precise values.}
\label{groupping}
\end{figure} 

\textit{Theory:} The second-order temporal coherence function allows
the investigation of photon grouping effects in terms of normally
ordered intensity-intensity correlations.  The normalized temporal
coherence function of a stationary radiation field consisting of a
single mode $\hat{a}$ is defined as the probability of a joint
detection of two photons at delayed times $t$ and $t+\tau$
\begin{equation}
g^{(2)}(\tau)=\frac{\langle \hat{a}^{\dagger}(t+\tau) \hat{a}^{\dagger}(t)
  \hat{a}(t+\tau) \hat{a}(t) \rangle}{\langle \hat{a}^{\dagger}(t) \hat{a}(t) \rangle^{2}}.
\label{g2_one_mode}
\end{equation}
For classical fields the operators are replaced by complex numbers,
and reordered such that at no delay we can write $g^{(2)}(0)=\langle
I^2\rangle / \langle I \rangle^2$ with $I$ the field intensity. This
quantity is bounded below by one, so classical fields cannot display
anti-bunching. The non-commutation of $\hat{a}$ and
$\hat{a}^{\dagger}$ prevents the quantum expression from being
reordered in this way, and allows for values of $g^{(2)}$ less than
one.  It is possible to rewrite Eq.~(\ref{g2_one_mode}) as a product
of intensities after a beamsplitter, such as measured on the HBT
interferometer
\begin{equation}
g^{(2)}(\tau)=\frac{\langle \hat{b}^{\dagger}(t+\tau)
\hat{b}(t+\tau)  \hat{c}^{\dagger}(t) \hat{c}(t) \rangle}{\langle
\hat{b}^{\dagger}(t) \hat{b}(t) \rangle  \langle
\hat{c}^{\dagger}(t) \hat{c}(t) \rangle}. \label{g2_two_mode}
\end{equation}
If $\hat{b}=(\hat{a}+\hat{v})/\sqrt{2}$ and
$\hat{c}=(\hat{a}-\hat{v})/ \sqrt{2}$ with $\hat{v}$ a vacuum mode,
then Eq.~(\ref{g2_one_mode}) and Eq.~(\ref{g2_two_mode}) are formally
equivalent.  The temporal coherence function witnesses the
anti-bunching effect when $g^{(2)}(0) < g^{(2)}(\tau\!\!\not =\!\!0)$.
The opposite inequality holds true for photon bunching.  This method
of measuring $g^{(2)}(\tau)$ is possible only when one has access to
the photon number correlations directly.

For experiments in the CV regime, a connection must be found for
measurements made via homodyne detection. A pair of homodyne detectors
will measure correlations between the field amplitude
$\hat{X}^{+}_a=\hat{a}+\hat{a}^{\dagger}$, or phase
$\hat{X}^{-}_a=-i(\hat{a}-\hat{a}^{\dagger})$, quadratures.  Therefore
we re-express the coherence function Eq.~(\ref{g2_two_mode}) in terms
of quadrature operators
\begin{equation}
g^{(2)}(\tau)=\frac{\sum_{i,j} \langle {\hat{X}_{b}^{i}}(t+\tau)^{2}
{\hat{X}_{c}^{j}}(t)^{2}\rangle \!-\! 2 \sum_{i,k} \langle
{\hat{X}_{k}^{i}}(t)^{2}\rangle \!+\! 4}{(\sum_{i} \langle
{\hat{X}_{b}^{i}}(t)^{2}\rangle-2)(\sum_{i} \langle
{\hat{X}_{c}^{i}}(t)^{2}\rangle-2)}
\label{quadratures}
\end{equation}
with $i,j=+,-$ and $k=b,c$. What makes this measurement technique
possible is that in the equation there are no cross-quadrature
correlation terms for a single mode. Hence each correlation or
variance term can be measured independently by recording the output of
the homodyne detectors (H1/H2 in Fig.~1), and a value for
$g^{(2)}(\tau)$ constructed accordingly.

We wish to consider the coherence function Eq.~(\ref{g2_two_mode}) for
a weakly squeezed and weakly displaced vacuum state. We work in the
Heisenberg picture. The operator $\hat{a}$ is first transformed
according to
\begin{equation}
\hat{D}^{\dagger}(\alpha)\hat{S}^{\dagger}(r) \hat{a}
\hat{S}(r)\hat{D}(\alpha)=\alpha + \hat{a} \cosh{r} -
\hat{a}^{\dagger} \sinh{r}
\label{transformation}
\end{equation}
where $\hat D$ and $\hat S$ are the unitary displacement and squeezing
operators respectively \cite{Walls&Milburn}, and we choose
$\alpha\!\in\!\!\Re$ and $r\!\in\!\!\Re$.  Then by using
Eq.~(\ref{transformation}) in Eq.~(\ref{g2_two_mode}) with $\tau=0$ we
find
\begin{equation}
g^{2}(0)=1+\frac{\sinh^{2}(r)\left( 2\alpha^{2}+\cosh(2r)-2\alpha^{2}\coth(r) \right)}
{\left( \alpha^{2}+\sinh^{2}(r)\right) ^{2}}.
\end{equation}
This function is plotted in Fig.~2 where bunching or anti-bunching
statistics can be found with the correct choice of $r$ and $\alpha$.
The stronger cases occur for states approaching the vacuum state
$\alpha\!=\!r\!=\!0$ for which $g^{(2)}(0)$ is undefined. Indeed it's
possible to approach the vacuum state while holding any value of
$g^{(2)}(0)$ constant.  The expression for $g^{(2)}(0)$ can be
extended to arbitrary $\tau$ by repeating the derivation but including
a delay of $\tau$ on mode $\hat b$ we obtain \begingroup
\setlength{\arraycolsep}{0pt}
\begin{eqnarray}
g^{(2)}(\tau) &=& \frac{1}{\left( \sinh^{2}(r)\! +\! {\alpha}^2
\right)^{\! 2}} \! \left \{\!
  \left( {\alpha}^2\!\! -\!\frac{1}{2}
  [\hat{a}(\tau),\hat{a}^{\dagger}(0)]\! \sinh(2r) \right)^2 \right.
\nonumber \\
&+& \left.
 \left(2 \!+\! [\hat{a}(0),\hat{a}^{\dagger}(\tau)]\! +\!
  [\hat{a}(\tau),\hat{a}^{\dagger}(0)]\right)\! {\alpha}^2\! \sinh^{2}(r)\right.
\nonumber \\
&+&  \left. \left(1\!+\!  [\hat{a}(0),\hat{a}^{\dagger}(\tau)]^2
\right) \!\sinh^{4}(r)
 \right \}.
 \label{pure_tau}
\end{eqnarray}
\endgroup Here, the commutation relations depend on the shape of the
filter used in the measurement process. The filter selects a
particular frequency mode given by
\begin{equation}
\hat{a}(\tau)\!\!=\!\!N^{-\frac{1}{2}}\!\! \int_{-\infty}^{\infty}
\hat{a}_{\omega} f(\omega) e^{i\tau \omega} d\omega
\end{equation}
where $f(\omega)$ is the filter function and
$N\!\!\!=\!\!\!\int_{-\infty}^{\infty} f(\omega)^2 d\omega$ is a
normalization factor.  Using $[\hat{a}_{\omega},
\hat{a}^{\dagger}_{\omega^{\prime}}] = \delta(\omega-\omega^{\prime})$
and choosing the filter $f(|\omega|\leqslant\Omega)\!\!=\!\!1$ and
zero elsewhere gives the commutation relations
$[\hat{a}(0),\hat{a}^{\dagger}(\tau)] =
[\hat{a}(\tau),\hat{a}^{\dagger}(0)] = \mathrm{sinc(\Omega \tau)}$.
Substituting these into Eq.~(\ref{pure_tau}) gives
\begin{eqnarray}
g^{(2)}(\tau) \!&=&\! \frac{1}{\left( \sinh^{2}(r)\!+ \!{\alpha}^2
\right)^2} \! \left \{\!
  \left( \!{\alpha}^2\!\! -\! \frac{1}{2} \mathrm{sinc(\Omega \tau)} \!\sinh(2r)
  \!\right)^2 \right.
\nonumber \\
&+& \left.
 2\left(1 +  \mathrm{sinc(\Omega \tau)}\right) {\alpha}^2 \sinh^{2}(r) \right.
\nonumber\\
&+& \left. (1+  \mathrm{sinc^2(\Omega \tau)}) \sinh^{4}(r) \right
\}.
\end{eqnarray}
This expression is specific to {\it pure} states, but can be
generalized for any Gaussian state having variances $V^{+}_{{\rm
    in}},V^{-}_{{\rm in}}$ and mean $\alpha_{{\rm in}}$ as measured at
the input to the HBT interferometer (see Fig.~1). The general form is
\begin{eqnarray}
g^{(2)}(\tau) &=&1+16 \mathrm{sinc(\Omega \tau)}\frac{(V^{+}_{\rm
in}\!-\!1)\alpha_{\rm in}^2}{(2-V^{+}_{\rm in}-V^{-}_{\rm in}-4
\alpha_{\rm
in}^2)^{2}}\nonumber \\
&+&\!2 \mathrm{sinc}^2(\Omega \tau)\frac{2\!+\!(V^{-}_{\rm
in}\!\!-\!2)V^{-}_{\rm in}\!\!+\!(V^{+}_{\rm in}\!\!-\!2)V^{+}_{\rm
in}}{(2-V^{+}_{\rm in}-V^{-}_{\rm in}-4 \alpha_{\rm in}^2)^{2}}.
\end{eqnarray}
This equation was used for making theoretical predictions that were
based on measurements $\{ V^{+}_{\rm in};V^{-}_{\rm in};\alpha_{\rm
  in} \}$. A comparison with the actual measured $g^{(2)}(\tau)$ could
then be made.

\textit{Experiment:} The schematic shown in Fig.~1 includes: the
source of squeezed light and displacement operation; the modified HBT
interferometer with homodyne detection and data acquisition. A Nd:YAG
laser emitted 1.5~W of coherent light at 1064~nm with a line-width of
1~kHz. A major fraction of this light was frequency-doubled and used
to pump an optical parametric amplifier (OPA) composed of a ${\rm
  MgO\!:\!LiNbO_{3}}$ crystal.  The OPA could be controlled so as to
de-amplify the transmitted seed beam by gain factors from 0.9 to 0.5.
This produced a bright beam $(\approx5~\mu\rm{W})$ having broadband
squeezing of the amplitude quadrature covering 3~MHz up to the cavity
line-width of 15~MHz. Corresponding levels of squeezing were measured
(at 6MHz) from $V^{+}_{\rm in}\!\!=\!0.89$ to $V^{+}_{\rm
  in}\!\!=\!0.55$, with values of purity $(V^{+}_{\rm in} V^{-}_{\rm
  in})$ ranging from $1.005$ to $1.18$, respectively. The displacement
operation was performed by injecting a bright amplitude-modulated
field at the 98:2 beam-splitter. The relative phase at the
beam-splitter was controlled, so that the quadrature of the squeezing
was preserved.

With the desired displaced squeezed state prepared, it was then
subjected to the second-order coherence measurement process as
performed via the modified HBT interferometer. The prepared state was
mixed with a vacuum state on a 50:50 beam-splitter. Each of the two
resulting output beams were sent on toward a balanced homodyne
detector, at whose core was a pair of matched photodiodes (ETX500)
with quantum efficiency estimated at $93\!\pm\!3\%$.  Care was taken
to keep the optical path lengths equal. The local-oscillators (LO)
were mode-matched into the test beam with $96\%$ visibility, and the
homodyne condition fulfilled by a signal/LO power ratio of $1/1000$.
The total detection efficiency was $86\%$. The LO phase was
controlled, so as to measure the required quadrature of either
amplitude or phase. The electronic signals of the photodetectors were
amplified, mixed-down at 6~MHz, and low-pass filtered, before being
over-sampled by a 12-bit digital-to-analogue converter at the rate of
240~kS/s. The time series data was then low-passed with a digital
top-hat filter and down-sampled to 120~kS/s, thereby ensuring a flat
power-spectrum with bandwidth 60~kHz. (This filter shape determines
the final $g^{(2)}(\tau)$ function). The variances and correlation
coefficients of the four permutations of the quadrature measurements
were calculated using approximately one million data points acquired
over 10 successive runs. These values were entered into
Eq.~(\ref{quadratures}) to get $g^{(2)}(\tau)$. The uncertainty in
each measurement ($68\%$ confidence interval) was calculated using
standard methods of error-analysis.

In this experiment we had the ability to freely vary the amount of
displacement, and also add broad-band noise to the amplitude
quadrature which simulated a thermal state. The squeezing strength and
purity were linked together by the optical loss and seed-coupling
mechanisms of the OPA, but some selection was possible by independent
variation of the pump and seed powers of the OPA. Additional optical
loss could be introduced by way of a variable optical-attenuator
placed before the HBT interferometer. Finally, the electronic signals
of each homodyne detector could be temporally shifted with respect to
one another, both before and after digital sampling.

\begin{figure}
  \center{\includegraphics[width=1\linewidth]{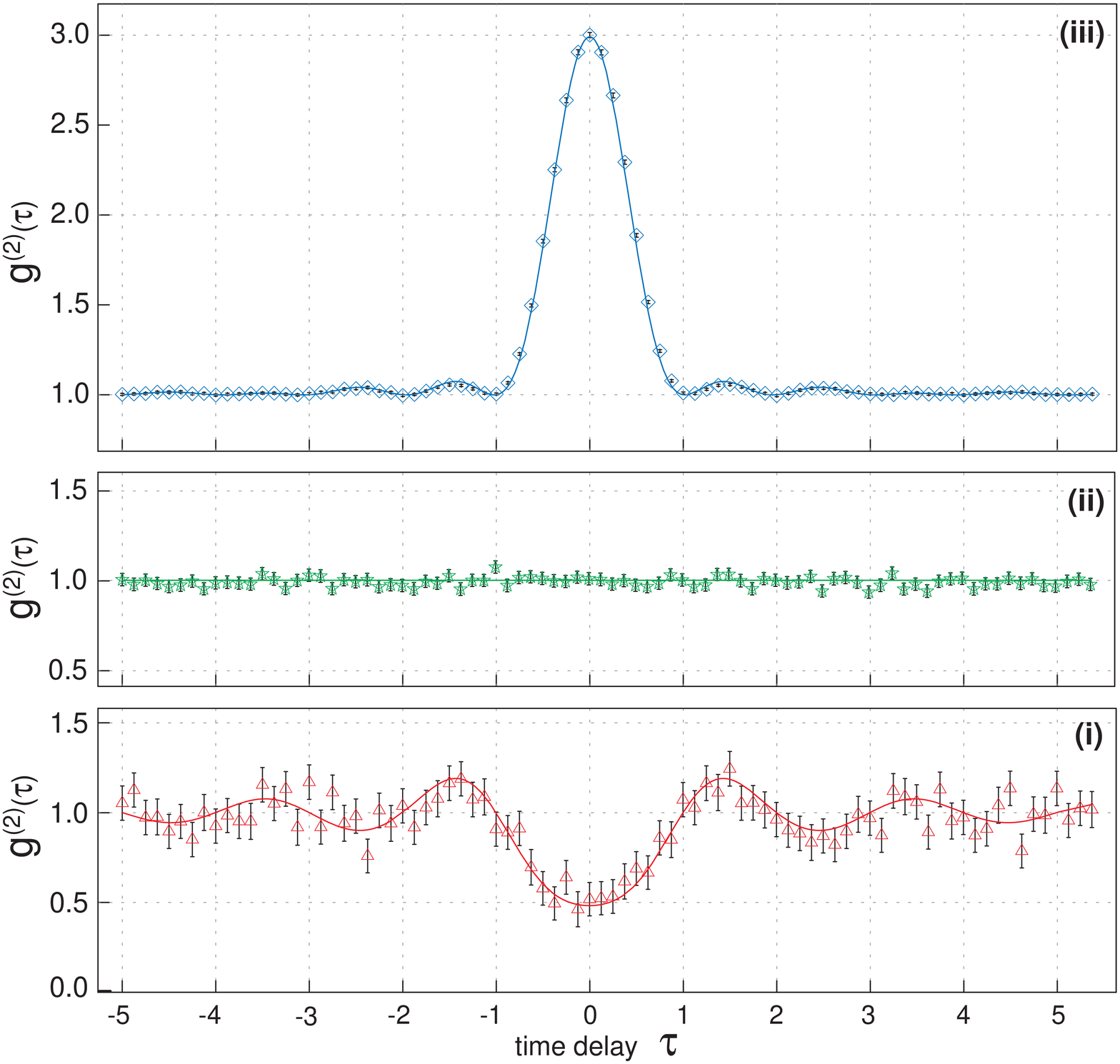}}
\caption{(Color online) Experimental measurement of 
  $g^{(2)}(\tau)$ with time delay $\tau$ in units of bandwidth
  $(\tau\Omega)$. (i) displaced squeezed state, (ii) coherent state,
  (iii) biased thermal state, curves are theoretical predictions.}
\end{figure} 

\begin{figure}
\center{\includegraphics[width=1\linewidth]{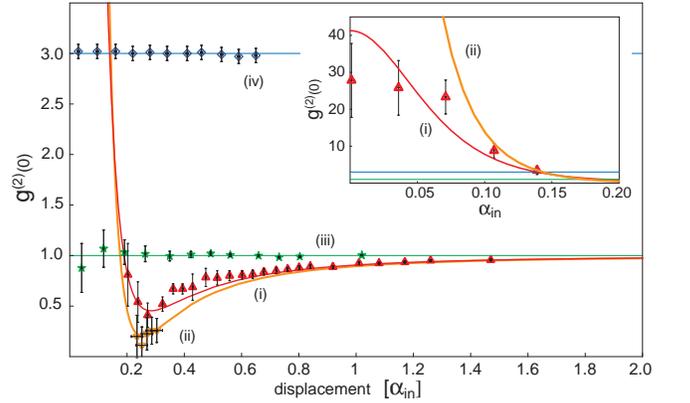}}
\caption{(Color online) Main figure and inset: Experimental measurement of 
  $g^{(2)}(0)$ as a function of displacement $\alpha_{\rm in}$. (i)
  displaced squeezed state, (ii) weak displaced squeezed state, (iii)
  coherent state, (iv) biased thermal state, curves are theoretical
  predictions.}
\end{figure} 

\textit{Results:} Photon anti-bunching statistics from a displaced
squeezed state were confirmed by the experimental results. Fig.~3(i)
shows how the measured value of $g^{(2)}(0)$ varied as a function of
time delay $\tau$. In the experiment, a squeezed state was prepared
and measured to have $\{ V^{+}_{\rm in}\!\!=\!0.902(1);V^{-}_{\rm
  in}\!\!=\!1.137(1) \}$. The state was then displaced by $\alpha_{\rm
  in}=0.257(1)$ which was the amount predicted to minimize $g^{(2)}$
for those particular variances. This prepared state yielded
$g^{(2)}(0)=0.44(22)$ for zero time delay. Then as $\tau$ was
increased, $g^{(2)}(\tau)$ approached unity. Thereby fulfilling the
requirement for photon anti-bunching $g^{(2)}(0)\!<\!g^{(2)}(\tau)$.

Displaced squeezed states present their most interesting behavior when
$g^{(2)}(0)$ is plotted along a range of displacements. Here, the
state can easily be manipulated into exhibiting either anti-bunching
or super-bunching statistics, depending only on the amount of
displacement applied. The experimental data is plotted in Fig.~4(i),
together with the theoretical curve as determined solely by $\{
V^{+}_{\rm in};V^{-}_{\rm in} \}$. Super-bunching statistics were
measured (see inset of Fig.~4) from the prepared state $\{ V^{+}_{\rm
  in}\!\!=\!0.901(3);V^{-}_{\rm in}\!\!=\!1.136(1) \}$ with zero
displacement $\alpha_{\rm in}\!\!=\!\!0.001(2)$ which produced
$g^{(2)}(0)\!=\!28(10)$.  On increasing $\alpha_{\rm in}$ still
further, the state first found a minimum value corresponding to
anti-bunching $g^{(2)}(0)=0.41(12)$ and then monotonically approached
one.

Aiming to observe even stronger anti-bunching, we prepared a squeezed
state approaching higher purity $\{ V^{+}_{\rm in}V^{-}_{\rm
  in}=0.890(2) \times 1.129(2) =1.005(3)\}$. This near pure state was
produced when the seed power to the OPA was reduced, which further
de-coupled extraneous noise sources. The $g^{(2)}(0)$ of this state
was measured for a range of displacements; Fig.~4(ii). The minimum
value was found at $\alpha_{\rm in}\!=\!0.252(2)$ which yielded
$g^{(2)}(0)\!=\!0.11(18)$. For comparison, a two-photon Fock state
would be limited to $g^{(2)}(0)\!\!=\!\!0.5$.

To fully characterize our HBT interferometer, we used two test states
that could readily be produced, and whose properties were well known;
the coherent state and the biased thermal-state (both with variable
displacement). Theory predicts that a coherent state will produce
$g^{(2)}(\tau)=1$ independent of $\alpha_{\rm in}$. This was confirmed
for $\tau=0$ in the measurement Fig.~4(iii), and also for variable
$\tau$ in Fig.~3(ii). Both sets of measurements kept within
$g^{(2)}(\tau)=1.00(6)$.

A biased thermal-state was made by taking a coherent state, and
applying to only the amplitude quadrature, a displacement driven by a
broad-band white-noise source. The state was called biased because $\{
V^{+}_{\rm in}\!\!\!>\!\!1;V^{-}_{\rm in}\!\!\!=\!\!1 \}$. The
prediction for a strongly biased thermal-state is $g^{(2)}(0)\!\simeq
\!3$ for $V^{+}_{\rm in}\!\!\gg\!\! 1$ and small displacments
$(\alpha_{\rm in}\!\!\sim\!\! 1)$. We prepared a state with $\{
V^{+}_{\rm in}\!\!\!=\!\!12.80(9);V^{-}_{\rm in}\!\!\!=\!\!1.039(1)
\}$ and varied $\alpha_{\rm in}$ from zero to $0.65(1)$ to obtain
Fig.~4(iv). The result was $g^{(2)}(0)\!=\!2.98(1)$ which adhered to
the theoretical prediction.  This state was also studied under
variable time delay; Fig.~3(iii).  At $\tau\!\!\!=\!\!\!0$, the
function was maximum $g^{(2)}(0)\!=\!2.98(1)$, and then fell towards
one with increasing $\tau$, according to the sinc function.  The
prepared state was $\{ V^{+}_{\rm in}\!\!=\!\!14.60(2);V^{-}_{\rm
  in}\!\!=\!\!1.025(8);\alpha_{\rm in}\!\!=\!\!0.258(1) \}$.  The
results from both the coherent state and biased-thermal state
measurements agreed with the theoretical predictions, and validated
our HBT interferometer.

A curious property of the $g^{(2)}$ measurement is that its value is
unchanged by prior mixing of the state with a vacuum state, hence
$g^{(2)}\!$ measurements are invariant to optical loss. We tested this
property by preparing a displaced-squeezed state $\{ V^{+}_{\rm
  in}\!\!\!=\!\!0.894(2);V^{-}_{\rm in}\!\!\!=\!\!1.139(2);\alpha_{\rm
  in}\!\!\!=\!\!0.255(2) \}$ and passing it through variable optical
attenuation. With zero attenuation, and total detection efficiency
$\eta_{\rm{det}}\!\!\!=\!\!\!86 \%$ we measured
$g^{(2)}(0)\!\!=\!\!0.67(16)$. Increasing the attenuation to
$\eta_{\rm{det}}\!\!\!=\!\!\!43 \%$ yielded
$g^{(2)}(0)\!\!=\!\!0.43(36)$. This showed that within a confidence
interval, $g^{(2)}\!$ was invariant to optical loss. The invariance
comes at the cost of increasing the uncertainty in the measurement. In
principle, the total measurement time can be increased to accommodate
an arbitrary level of attenuation.

In summary, we have used a continuous-variable measurement scheme to
experimentally probe the second-order coherence function of quantum
states of light. The scheme was based on the Hanbury-Brown Twiss
intensity interferometer, but used homodyne detection of the
quadrature amplitudes, in contrast to discrete-variable single photon
counting. By preparing displaced squeezed states with the appropriate
parameters, we were able to observe photon anti-bunching as witnessed
by $g^{(2)}(0)\!\!=\!\!0.11(18)$, which is a non-classical effect, and
a direct manifestation of the quantum nature of light.

This work was supported by the Australian Research Council Discovery
Grant scheme, and by the MEN Grant No.~1~PO3B~137~30.


\begin{thebibliography}{99}
  
\bibitem{HBT} R.~Hanbury~Brown, and R.~Q.~Twiss, {\it Nature} {\bf
    177}, 27, (1956).
  
\bibitem{Condensed1} M.~Henny, S.~Oberholzer, C.~Strunk, T.~Heinzl,
  K.~Ensslin, M.~Holland, and C.~Sch\"onenberger, {\it Science} {\bf
    284}, 296 (1999); W.~D.~Oliver, J.~Kim, and R.~C.~Liu,
  Y.~Yamamoto, {\it Science} {\bf 284}, 299 (1999).
  
\bibitem{Atomic} M.~Yasuda, and F.~Shimizu, {\it \prl} {\bf 77}, 3090
  (1996).
  
\bibitem{Walls&Milburn} D.~F. Walls and G.~J. Milburn, \emph{Quantum
    Optics} (Cambridge University Press, 1994).
  
\bibitem{Astronomy} R.~Hanbury~Brown, and R.~Q.~Twiss, {\it Nature}
  {\bf 178}, 1046 (1956).
  
\bibitem{Particles} G.~Goldhaber, S.~Goldhaber, W.~Lee, and A.~Pais,
  {\it Phys. Rev.} {\bf 120}, 300 (1960); D.~H.~Boal, C.-K.~Gelbke,
  and B.~K.~Jennings, {\it Rev. Mod. Phys.} {\bf 62}, 553 (1990).
  
\bibitem{Glauber} R.~Glauber, {\it Phys. Rev.} {\bf130}, 2529 (1963).
  
\bibitem{Kimble} H.~J.~Kimble, M.~Dagenais, and L.~Mandel, {\it \prl}
  {\bf 39}, 691 (1977); F.~Diedrich and H.~Walther, {\it \prl} {\bf
    58}, 203 (1987); G.~Leuchs, and U.~L.~Andersen, {\it Laser
    Physics} {\bf 15}, 129 (2005).
  
\bibitem{RarityAndNogueira} J.~G.~Rarity, P.~R.~Tapster, and
  E.~Jakeman, {\it Opt. Commun.} {\bf 62}, 201 (1987);
  W.~A.~T.~Nogueira, S.~P.~Walborn, S.~P\'adua, and C.~H.~Monken, {\it
    \prl} {\bf 86}, 4009 (2001).
  
\bibitem{Koashi} M.~Koashi, K.~Kono, T.~Hirano, and M.~Matsuoka, {\it
    \prl} {\bf 71}, 1164 (1993).
  
\bibitem{QDots} P.~Michler, A.~Imamo\u{g}lu, M.~D.~Mason,
  P.~J.~Carson, G.~F.~Strouse, and S.~K.~Buratto, {\it Nature} {\bf
    406}, 968 (2000); C.~Santori, D.~Fattal, J.~Vu\v{c}kovi\'{c},
  G.~S.~Solomon, Y.~Yamamoto, {\it Nature} {\bf 419}, 594 (2002).
  
\bibitem{SinglePhoton} B.~Lounis, and W.~E.~Moerner, {\it Nature} {\bf
    407}, 491 (2000); B.~Darqui\'e, M.~P.~A.~Jones, J.~Dingjan,
  J.~Beugnon, S.~Bergamini, Y.~Sortais, G.~Messin, A.~Browaeys, and
  P.~Grangier, {\it Science} {\bf 309}, 454 (2005).
  
\bibitem{Schellekens} M.~Schellekens, R.~Hoppeler, A.~Perrin,
  J.~V.~Gomes, D.~Boiron, A.~Aspect, and C.~I.~Westbrook, {\it
    Science} {\bf 310}, 648 (2005).
  
\bibitem{Oettl} A.~\"{O}ttl, S.~Ritter, M.~K\"ohl, and T.~Esslinger,
  {\it \prl} {\bf 95}, 090404 (2005).
  
\bibitem{McAlisterAndWebb} D.~F.~McAlister, and M.~G.~Raymer, {\it
    \pra} {\bf 55}, R1609 (1997); J.~G.~Webb, T.~C.~Ralph, and
  E.~H.~Huntington, {\it \pra} {\bf 73}, 033808 (2006).
  
\bibitem{StolerAndMahran} D.~Stoler, {\it \prl} {\bf 33}, 1397 (1974);
  M.~H.~Mahran, and M.~V.~Satyanarayana, {\it \pra} {\bf 34}, 640
  (1986).

\end{thebibliography}
\end{document}